\documentclass[acmlarge]{acmart}
\AtBeginDocument{%
  \providecommand\BibTeX{{%
    \normalfont B\kern-0.5em{\scshape i\kern-0.25em b}\kern-0.8em\TeX}}}
\makeatletter

\newcommand{\Rmnum}[1]{\expandafter\@slowromancap\romannumeral #1@}
\makeatother

\usepackage{CJK}
\usepackage{textcomp}
\usepackage{colortbl}
\usepackage{graphicx}
\usepackage{multirow}
\usepackage{algorithmic}
\usepackage{algorithm}

\copyrightyear{2024}
\acmYear{2024}
\setcopyright{acmlicensed}\acmConference[CHI '24]{Proceedings of the CHI Conference on Human Factors in Computing Systems}{May 11--16, 2024}{Honolulu, HI, USA}
\acmBooktitle{Proceedings of the CHI Conference on Human Factors in Computing Systems (CHI '24), May 11--16, 2024, Honolulu, HI, USA}
\acmDOI{10.1145/3613904.3642912}
\acmISBN{979-8-4007-0330-0/24/05}

\acmSubmissionID{7908}

\begin{document}

\title{To Reach the Unreachable: Exploring the Potential of VR Hand Redirection for Upper Limb Rehabilitation}

\author{Peixuan Xiong}
\affiliation{%
  \institution{The Hong Kong University of Science and Technology (Guangzhou)}
  \city{Guangzhou}
  \country{China}}
\email{pxiong843@connect.hkust-gz.edu.cn}

\author{Yukai Zhang}
\affiliation{%
  \institution{The Hong Kong University of Science and Technology (Guangzhou)}
  \city{Guangzhou}
  \country{China}}
\email{yzhang118@connect.hkust-gz.edu.cn}

\author{Nandi Zhang}
\affiliation{%
  \institution{University of Calgary}
  \city{Calgary}
  \country{Canada}}
\email{nandi.zhang@ucalgary.ca}

\author{Shihan Fu}
\affiliation{%
  \institution{The Hong Kong University of Science and Technology (Guangzhou)}
  \city{Guangzhou}
  \country{China}}
\email{sfu663@connect.hkust-gz.edu.cn}

\author{Xin Li}
\affiliation{%
  \institution{The Third Affiliated Hospital}  \institution{Sun Yat-sen University}
  \city{Guangzhou}
  \country{China}}
\email{lixin36@mail.sysu.edu.cn}

\author{Yadan Zheng}
\affiliation{%
  \institution{The Third Affiliated Hospital} \institution{Sun Yat-sen University}
  \city{Guangzhou}
  \country{China}}
\email{zhengyad@mail.sysu.edu.cn}

\author{Jinni Zhou}
{\footnotesize
\affiliation{%
  \institution{The Hong Kong University of Science and Technology (Guangzhou)}
  \city{Guangzhou}
  \country{China}}
\email{eejinni@hkust-gz.edu.cn}
}
\author{Xiquan Hu}
\authornote{corresponding author}
\affiliation{%
  \institution{The Third Affiliated Hospital} \institution{Sun Yat-sen University}
  \city{Guangzhou}
  \country{China}}
\email{huxiquan@mail.sysu.edu.cn}

\author{Mingming Fan}
\authornotemark[1]
\affiliation{%
%\institution{Computational Media and Arts Thrust} 
\institution{The Hong Kong University of Science and Technology (Guangzhou)}
  \city{Guangzhou}
  \country{China}
 }
 \affiliation{
 %  \institution{Division of Integrative Systems and Design \& Department of Computer Science and Engineering}
\institution{The Hong Kong University of Science and Technology}
  \city{Hong Kong SAR}
  \country{China}
}  
\email{mingmingfan@ust.hk}

\renewcommand{\shortauthors}{Peixuan Xiong, et al.}

\begin{abstract}
Rehabilitation therapies are widely employed to assist people with motor impairments in regaining control over their affected body parts. Nevertheless, factors such as fatigue and low self-efficacy can hinder patient compliance during extensive rehabilitation processes. Utilizing hand redirection in virtual reality (VR) enables patients to accomplish seemingly more challenging tasks, thereby bolstering their motivation and confidence. While previous research has investigated user experience and hand redirection among able-bodied people, its effects on motor-impaired people remain unexplored. In this paper, we present a VR rehabilitation application that harnesses hand redirection. Through a user study and semi-structured interviews, we examine the impact of hand redirection on the rehabilitation experiences of people with motor impairments and its potential to enhance their motivation for upper limb rehabilitation. Our findings suggest that patients are not sensitive to hand movement inconsistency, and the majority express interest in incorporating hand redirection into future long-term VR rehabilitation programs.
\end{abstract}




%\received{2 September 2023}
%\received[revised]{12 November 2023}
%\received[accepted]{5 January 2024}

\maketitle
\section{INTRODUCTION}
\begin{figure*}[!ht]
\centering
\includegraphics[width=0.89\linewidth]{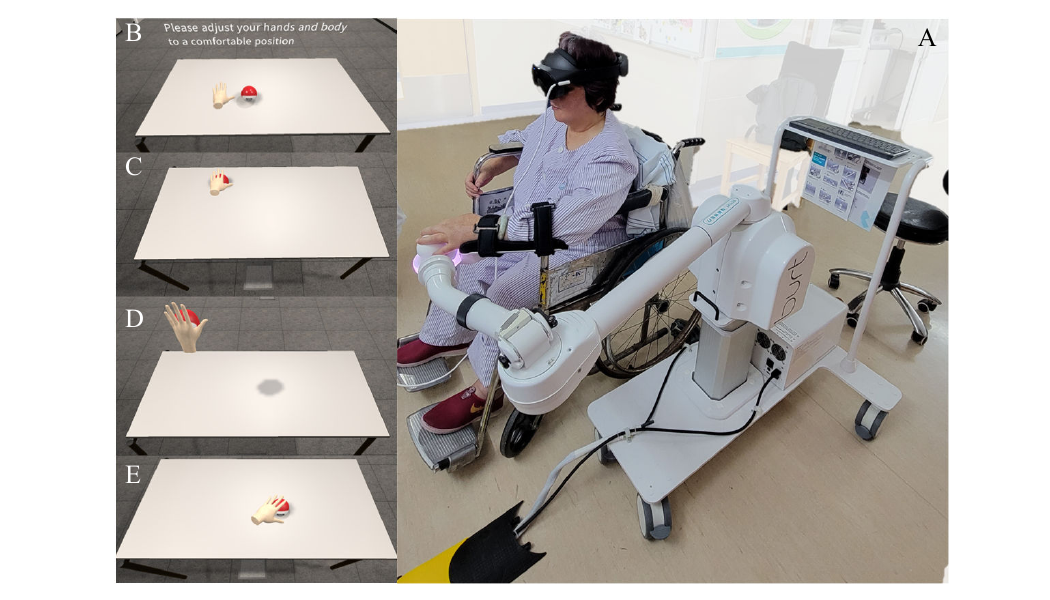}
\caption{Overview of the study: (A) The participant wears the BURT and the VR headset. BURT helps participants cancel out the burden of gravity on the affected arm. (B) Warm-up for VR Rehabilitation scene with text hint: ``Please adjust your hands and body to a comfortable position'' (C) The participant performs reaching in VR, corresponding to depth; (D) The participant performs shoulder flexion in VR, corresponding to vertical; (E) The participant performs horizontal abduction in VR, corresponding to horizontal.}
\Description{The figure given a demonstration of user study setup and VR scenarios: (A) The participant is sitting in a wheelchair, with left hand fixed on the BURT rehabilitation robot. The robot is a tandem robot and ends with a spherical handle, while the hand is placed on the platform of the penultimate segment. An Oculus Quest Pro VR headset is equipped. (B) This is the warm-up VR scene, with a red ball and a left hand on the table. A sentence informing participants to adjust their hands is hovering on the table, facing the participant. (C) The red ball is placed deeper. The participant tries to reach it. (D) The red ball is placed higher above the table, and the participant tries to raise his or her hand to reach it. (E) The red ball is placed far to the training side on the table, and the participant uses horizontal abduction to reach it.}
\label{apparauts}
\end{figure*}

Motor impairments, especially hand and arm mobility impairments, are common symptoms among people who live with cerebral palsy (CP), amyotrophic lateral sclerosis (ALS), spinal cord injury (SCI), and brain stroke. Such motor impairments negatively affect their independent living, including performing activities of daily life \cite{hatem2016rehabilitation,zhao2019clinical}. Rehabilitation therapies are commonly adopted to help people with motor impairment regain strength and control over their impaired body parts and may help them fully or partially restore their motor ability  \cite{chu2018robot}. 
Performing rehabilitation exercises in a hospital is a common step that many people with motor impairments (e.g., stroke survivors) take when following rehabilitation therapies. Although there are some methods that allow patients to exercise for rehabilitation without medical equipment, such as games \cite{gorvsivc2019pilot, gorsic2016design}, haptic exercises \cite{9671087}, or 
% You: TODO: rephrase
% Sep 11, 2023 9:07 PM • Edit
% Hit Enter to reply
performing exercises following the initial demonstrations from physiotherapists \cite{lam2014improving}. These methods still cause frustration, tediousness, and lack of motivation \cite{palaniappan2018developing} because the patients can not perceive their improvement directly. 
People with motor impairment receiving more physiotherapy treatment recover faster \cite{partridge2000dosage,glasgow2004can}. The majority of rehabilitation hospitals have widely adopted constraint-induced movement therapy (CIMT) for patients, which requires intensive, repetitive daily activities of the affected side while constraining the use of the less-affected side. This approach has demonstrated effective therapeutic outcomes \cite{kwakkel2007constraint,nijland2011constraint}. 
However, the high intensity of this treatment - demanding more than six hours of daily exercise, five days a week, for a duration exceeding two weeks - also leads to fatigue and reduced compliance for patients \cite{lin2007effects,kim2018effects}. 
% On one hand, the reduced compliance due to fatigue results in a decreased willingness of patients to engage in rehabilitation activities. 
On the other hand, while substantial training efforts might not yield noticeable improvements, patients remain highly sensitive to their failures when employing the affected limbs, discouraging them from engaging in the exercise. As a result, it is important to motivate patients during the lengthy rehabilitation process by allowing them to \textit{perceive improvement} to ensure effective rehabilitation.

Recent developments in Virtual Reality (VR) have shown great potential for making rehabilitation more engaging. VR-based rehabilitation allows patients to carry out rehabilitative exercises while immersing themselves in an interactive virtual environment, increasing the level of engagement which can lead to better therapeutic outcomes \cite{Yeh2017,tailorvr, Merians2002, Lewis2012, Zimmerli2013-qo, Cano_Porras2018-qf}. A promising technique for rehabilitation is \textit{hand redirection}, which breaks the exact one-to-one mapping (i.e. completely synchronized movement) between a patient's physical hands and the corresponding virtual hands in VR by modifying the visual movements of the virtual hands. With hand redirection, the hand movements of patients with motor impairment can be increased such that they can accomplish seemingly more difficult tasks. This has the potential value to increase their extrinsic motivation. However, such increased movement can potentially reduce users' sense of hand ownership, which is the feeling that the hand belongs to oneself \cite{blanke2012multisensory}. The sense of hand ownership can be measured by the noticeability of hand redirection. A reduced sense of hand ownership makes hand redirection more noticeable, which may negatively impact rehabilitation as it reminds patients that their good performance results from external help, thereby reducing their motivation \cite{wenk2022hiding}. As a result, it is crucial to find an optimal form of hand redirection that can facilitate improved performance in patients while maintaining a strong sense of hand ownership.

Previous studies on hand redirection, however, are primarily conducted with able-bodied users  \cite{zenner2019estimating,li2022modeling}, while little is known about how people with motor impairments perceive hand redirection in VR and how different hand redirection techniques might improve their motivation in rehabilitation. 
% \revision{Moreover, blindly reducing the difficulty of rehabilitation would have the opposite effect, as patients would not be able to obtain sufficient exercise. Therefore, we also intend to explore the design of new hand redirection methods that would allow patients to feel a real sense of progress while still maintaining a certain level
% of difficulty. }
Previous works have explored two primary paradigms: Pre-offset (applying offsets at the initiation of a motion\cite{Han2018,Benda2020}, Fig.2 B) or scaling offset (gradually throughout the entire hand movement\cite{kohli2013redirected,Cheng2017,Han2018,Esmaeili2020,zenner2019estimating}, Fig.2 C). In addition to these techniques, we propose a new hand redirection method that applies offsets gradually as the hand approaches the target.

Overall, our goal was to increase patients’ motivation and effort in rehabilitation while retaining the sense of hand ownership.
In this paper, we conducted a user study with 11 patients with upper limb impairments at a local hospital to understand its effect on motivation in comparison with no hand redirection and other hand redirection techniques in the literature. Thus, our research question (RQ) is: \textbf{How do different hand redirection techniques affect the effort, motivation, and sense of hand ownership of patients in VR-based rehabilitation?} Our findings show that participants hardly noticed hand redirection and they showed more effort and motivation in VR-based rehabilitation with hand redirection. 
In summary, we made the following contributions:
\begin{itemize}
\item[$\bullet$] We proposed a novel hand redirection technique that retains the sense of hand ownership while improving participant's training effort and motivation.
\item[$\bullet$] We conducted an empirical study with upper-limb-impaired patients through a VR rehabilitation application that harnesses hand redirection.
\end{itemize}

\section{BACKGROUND AND RELATED WORK}

\subsection{Rehabilitation for Upper-limb Motor Impairments}
% In the past decade, many rehabilitation methods have been developed around the motor learning paradigm targeting upper-limb motor impairments. Conventional methods including range of motion (ROM) exercises, strengthening exercises, constraint-induced movement therapy (CIMT), task-based training, etc. have been extensively studied. Restoration of motor functions requires repetitive, intense training.  functional electrical stimulation (FES), task-based training, mirror therapy, brain-computer interfaces (BCI), robot-assisted therapy (RAT), VR-based interventions, video game rehabilitation
Upper-limb motor impairment is a prominent concern due to its global prevalence and disabling effects. Prioritizing recovery empowers patients for independent daily activities, reducing reliance on caregivers and professionals, and enhancing overall quality of life ~\cite{dijkers1997quality}.

Various therapeutic approaches have been explored for stroke patients, including robot-assisted training ~\cite{blank2014current, Colombo2007Design, Frisoli2009-dq}, motor training sessions~\cite{Dietz2002}, constraint-induced movement therapy (CIMT) \cite{kwakkel2007constraint,Etoom2016-vk,Reiss2012-xz,Fabbrini2014-ls}, and physical therapy~\cite{healthcare10020190,Levin2021-xj,hatem2016rehabilitation}. For instance, Colombo et al.~\cite{Colombo2007Design} introduced motivating robotic devices for upper limb rehabilitation, incorporating gaming elements (challenges, rewards, rules. etc) to enhance patient engagement and emphasizing regular performance evaluation.

Motivation plays a crucial role in driving improved performance in rehabilitation and is a critical determinant of outcomes ~\cite{Verrienti2023-sn}. Active participation reflects motivation, while passivity indicates a lack thereof ~\cite{Concept2002}. Traditional methods often struggle to maintain this balance, constrained by the need to moderate training difficulty in light of patient frustration levels \cite{Spindler2019-yr}. In this work, we tackle this issue by leveraging personalized parameters that are tailored to each participant.

\subsection{VR for Rehabilitation}
Rehabilitating motor-impaired patients aims at promoting effective motor learning~\cite{Kiper2018-gi}, a task often hindered by the challenges of intensive training and motivation diminished by low self-efficacy. VR offers a solution, proving effective across conditions including cerebral palsy~\cite{Qiu2009-yc,Metin_Okmen2019-hk,Liu2022}, Parkinson's~\cite{Powell2014,Lei2019-zw,Winter2021-qj}, stroke~\cite{tailorvr, Elor2018,Huang2019-tq,Salisbury2020,Tannus2022,Duval2022}, and injuries~\cite{pietrzak2014using,maresca2018novel,ustinova2014virtual}.

Prior studies have explored the effect of VR in rehabilitation of enhancing patient engagement through visual and auditory feedback on movement~\cite{Yeh2017,tailorvr, Merians2002,Lewis2012,Zimmerli2013-qo,Cano_Porras2018-qf}. Zimmerli et al. highlighted the pivotal role of interactivity  (ie, functional feedback to motor performance) of VR in engaging patients in rehabilitation~\cite{Zimmerli2013-qo}. Other studies showcase the effectiveness of combining VR with other approaches, such as conventional physical therapy \cite{Rodriguez-Hernandez2023-rs,Nair2022-ld,Oh2019-ni,Hsu2022-rd,Jerome-Christian2012}, robotic-assisted rehabilitation~\cite{EvaluateVRrobotics,Zanatta2023-dw,Qiu2009-yc}, and exoskeletons~\cite{Frisoli2009-dq,Thielbar2014-vz,Klamroth-Marganska2014-gi} in improving motor functions.

VR has been widely explored as a rehabilitation interface to promote patient motivation in different ways. For instance, Dias et al.'s VR minigames, incorporating competitive elements and encouraging sounds, exemplify strategies to enhance motivation and prevent frustration during rehabilitation~\cite{Dias2019}. While prior VR interventions often focus on creating motivating content, the potential of leveraging VR's customizability to tackle the self-efficacy-difficulty challenge encountered in real-world rehabilitation remains underexplored. This paper aims to address this gap in the literature.
%%%这里好像没说清楚前文具体做了什么导致不足，而要引出我们的贡献。
%When playing rehabilitation games at the hospital, most of the therapy are in public sight. The patients have no private. However the VR headset can enhance the privacy
\subsection{Hand Redirection in VR}
% citations not fully added
Hand redirection techniques in VR have emerged as an illusion-based method to manipulate the mapping of real-to-virtual hand positions, thereby enhancing user experiences and interactions. 

An essential consideration in the implementation of hand redirection techniques is to ensure that users maintain a strong sense of ownership in the virtual environment \cite{noticeability}. Prior studies have demonstrated that inducing a body ownership illusion during VR rehabilitation can improve the therapeutic effect over the non-embodied rehabilitation in VR \cite{perez2017increasing,cha2021novel,fregna2022novel}. As such, a key focus in prior research has been on determining the detection thresholds of these offset across various motion axes \cite{Esmaeili2020,Hartfill2021, noticeability}. While previous research has explored people's ability to perceive hand redirection in VR environments \cite{zenner2019estimating}, no prior studies have specifically addressed patients with upper limb impairments. Given that individuals in this patient group typically exhibit lower cognitive abilities, we hypothesized that their detection threshold (i.e., just noticeable difference (JND)) of visual-proprioceptive conflict might be higher than that of healthy individuals.
To address this, our research aims to systematically evaluate various hand redirection techniques, examining their impact on patient motivation and the preservation of ownership in VR.

\section{USER STUDY}
% \label{Method}
To understand how different hand redirection techniques affect the motivation, effort, and sense of hand ownership of people with motor impairments during rehabilitation, we conducted an empirical user study at a local hospital with 11 participants. A VR-based rehabilitation exercise application with three different hand redirection techniques (pre-offset, scaling offset, and post-offset) is implemented. During the study, patients with upper limb impairment are instructed to reach targets in VR at various positions using their affected arm with and without hand redirection because reaching is the most common movement in daily activities and serves as the fundamental movement in stroke rehabilitation training~\cite{alt2015kinematic}. 

\subsection{Hand Redirection Techniques}
\begin{figure*}[!h]
\centering
\includegraphics[width=\linewidth]{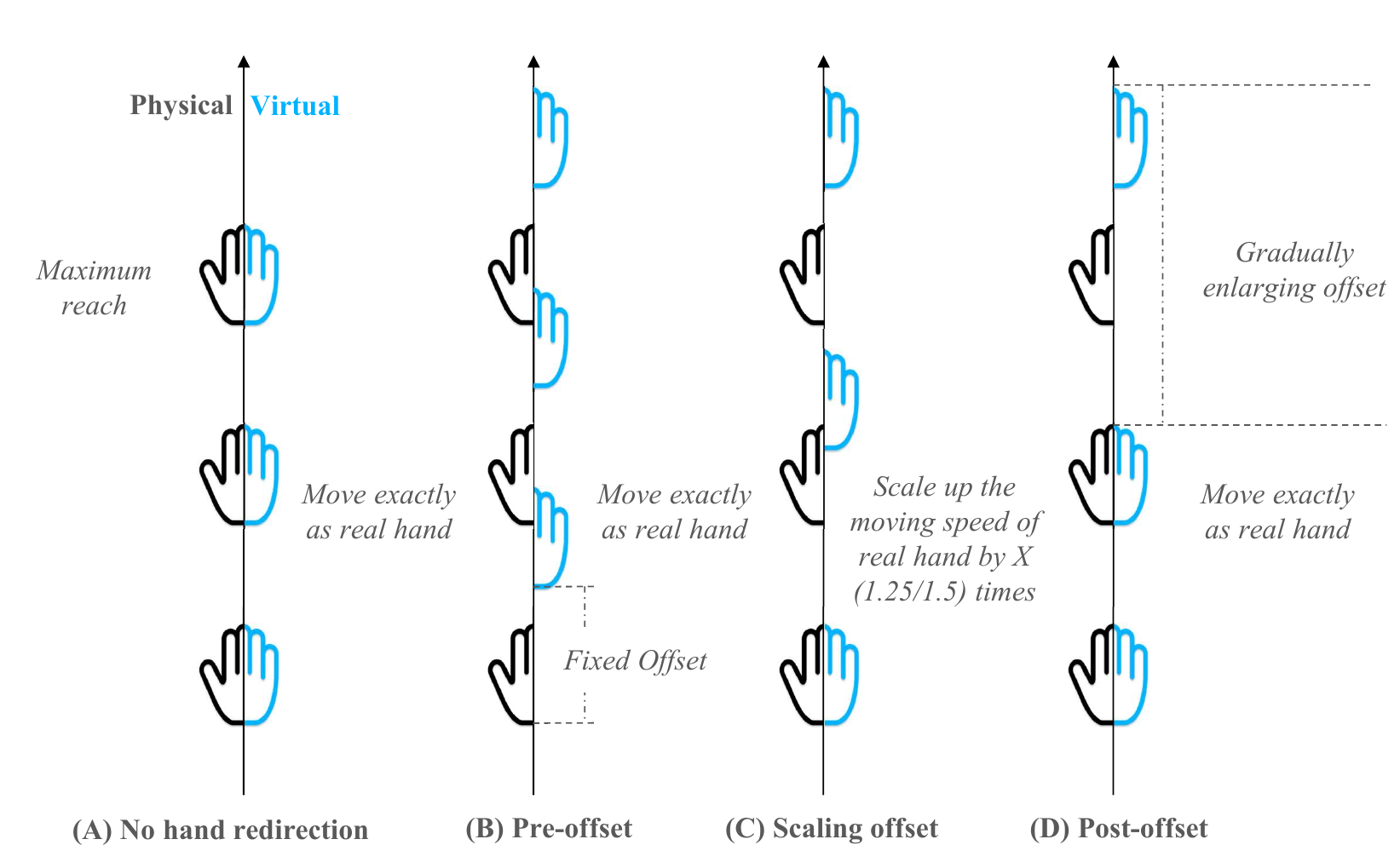}
\caption{Hand redirection techniques: (A) No hand redirection; (B) Pre-offset: the virtual hand is displaced a fixed offset from the real hand from the beginning to the end of hand motion; (C) Scaling offset: the virtual hand scales the movement of the real hand by a fixed factor; (D) Post-offset: the virtual hand gradually enlarging the movement when the real hand approaches its maximum reach. In all offset conditions, the parameters are set such that the virtual hand offsets the real hand the same distance when the real hand is at the maximum reach of the user.}
\Description{This figure includes three hand redirection techniques and a baseline. Pictures of the hand indicate the movement and distinguish the real and virtual hand. In all conditions, there is a vertical line dividing the physical hand and the virtual hand. The real hand is black and above, and the virtual hand is blue and below. The virtual hands are from semi-transparent to solid, with the direction of the arrow from left to right. In all offset conditions, the parameters are set such that the virtual hand offsets the real hand the same distance when the real hand is at the maximum reach of the user. (A) This is no hand redirection baseline condition. The movement of real and virtual hands is the same in this condition. (B) This is Pre-offset condition, and the virtual hand is displaced a fixed offset from the real hand from the beginning to the end of hand motion; (C) This is the Scaling offset condition. The virtual hand scales the movement of the real hand by a fixed factor, so its speed is higher than the real hand; (D) This is the Post-offset condition. The virtual hand drifts toward the target when the real hand approaches its maximum reach.}
\label{techniques}
\end{figure*}
Three different hand redirection techniques were tested in the study: Pre-offset, Scaling offset, and Post-offset, as illustrated in Fig.  \ref{techniques}. 

\textbf{Pre-offset.} The Pre-offset technique offsets the virtual hand from the physical hand by a constant displacement towards the target regardless of the real hand position. By bringing the affected hand closer to the target, the Pre-offset technique makes it easier for patients to succeed in the target-reaching task, potentially increasing their motivation to engage in the exercise. 

\textbf{Scaling offset.}  The Scaling offset technique scales the position of the virtual hand based on the real hand position from an original position. With a scaling factor larger than 1, it enlarges the virtual hand movement, making it easier for patients to reach the target. Similar to the Pre-offset, this might encourage the patient at the cost of lower training efforts.

However, as the task becomes easier, this may come at the cost of reduced training efforts, harming the therapeutic effect \cite{takebayashi2022impact}. To resolve this problem, we proposed a new hand redirection technique.

\textbf{Post-offset.}   Unlike Pre-offset and Scaling-offset, the displacement between the virtual and physical hand for Post-offset is only applied when the physical hand approaches its maximum reach. As the physical hand reaches a threshold close to its maximum reach, a gradually enlarging offset is applied, extending the virtual hand toward the target until the offset reaches a preset constant displacement. By applying the offset only when the hand approaches its limit, the Post-offset has the potential to induce the needed workload for the desirable therapeutic effect while maintaining a high success rate in accomplishing the tasks, keeping patients motivated.  
% The formula can be expressed as follows:
% \begin{algorithm}[!h]
%     \caption{Application program of Post-offset}
%     \label{alg:AOP}
%     \renewcommand{\algorithmicrequire}{\textbf{Input:}}
%     \renewcommand{\algorithmicensure}{\textbf{Output:}}
    
%     \begin{algorithmic}[1]
%         \REQUIRE real hand position $\vec{p}_{r}$, initial position $\vec{p}_{ini}$, max position $\vec{p}_{max}$, factor of threshold $t$, factor of the coefficient $g$   %%input
%         \ENSURE virtual hand position $\vec{p}_{v}$    %%output
        
%         \IF {$\left (  \vec{p}_{r}   > t\times \vec{p}_{max} \right )  $}
%             \STATE  $\vec{p_{v}} = p_{r}$, execute in 2 seconds
%         \ENDIF 
        
%         \WHILE{$\left (  \vec{p}_{r}   > t\times \vec{p}_{max} \right )  $}
%             \STATE $\vec{p}_{v}=\vec{p}_{r}  +  \left (  g- 1\right ) \times \left ( \vec{p}_{max} -\vec{p}_{ini}\right ) $
%         \ENDWHILE

%         \RETURN $\vec{p}^{v}$
%     \end{algorithmic}
% \end{algorithm}

% where $p_{v}$ is the the virtual hand position, $p^{r}$ is the position of the physical hand. $p_{max}$ is the participants' maximum reaching position, and $p^{ini}$ is the participants' initial hand position at the beginning of the trial. $g$ is the coefficient, and $t$ is the threshold, which is 0.8 in this work. Speed in each frame is the same. The offset will execute in two seconds after the hand is detected beyond the threshold.

\subsection{Apparatus}
We conducted the user study in the operational therapy room at a local hospital. We used the Barrett Upper-extremity Robotic Trainer (BURT) to support the patient's affected arm in order to relieve them from the burden of gravity so that more patients with low hand mobility could be included in the study. 
We conducted the study using the Oculus Quest Pro and Oculus Quest 2 headset, powered by an Intel Core i7 CPU and an NVIDIA GeForce RTX 3070ti GPU. The VR-based rehabilitation application was developed with Unity. With hand tracking, participants can see the virtual representation of their real hands in the VR scene. 
\subsection{Participants}
\begin{table*} [!h]
    \centering 
    \caption{Participants' demographic information and their symptoms} 
\Description{The first column is ID, containing a unique identifier for each participant, ranging from 1 to 11. The second column provides the age of each participant. The third column is Affect side. The fourth column is the symptom, which describes the underlying medical condition or injury of each participant. Participant 1 is a 53-year-old individual with left-side affect. This participant has moderate motor impairments, with the ability to move against gravity less than 50\%.
Participant 2 is a 56-year-old individual with left-side affect. This participant has moderate motor impairments, with the ability to move against gravity ranging from 50\% to 100\%.
Participant 3 is a 51-year-old individual with left-side affect. This participant has moderate motor impairments, with the ability to move against gravity ranging from 50\% to 100\%.
Participant 4 is a 51-year-old individual with left-side affect. This participant has mild motor impairments, with the ability to move against gravity and resistance ranging from 50\% to 100\%.
Participant 5 is a 66-year-old individual with left-side affect. This participant has moderate motor impairments, with the ability to move against gravity ranging from 50\% to 100\%.
Participant 6 is a 66-year-old individual with right-side affect. This participant has mild motor impairments, with the ability to move against gravity and resistance ranging from 50\% to 100\%.
Participant 7 is a 52-year-old individual with right-side affect. This participant has moderate motor impairments, with the ability to move against gravity ranging from 50\% to 100\%.
Participant 8 is a 60-year-old individual with right-side affect. This participant has mild motor impairments, with the ability to move against gravity and resistance less than 50\%.
Participant 9 is a 22-year-old individual with left-side affect. This participant has severe motor impairments, with the ability to move without gravity ranging from 50\% to 100\%.
Participant 10 is a 29-year-old individual with left-side affect. This participant has severe motor impairments, with the ability to move without gravity less than 50\%.
Participant 11 is a 50-year-old individual with left-side affect. This participant has severe motor impairments, with the ability to move without gravity less than 50\%.
}
      \begin{tabular}{cccc}
\hline
ID & age & Affect side & Motor impairments \\
% less than 50\% 
\hline
1 & 53 & Left & Moderate, able to move the arm against gravity (Grade 3 in MMT) \\
2 & 56&  Left & Moderate, able to move the arm against gravity (Grade 3 in MMT) \\
3 & 51 & Left & Moderate, able to move the arm against gravity (Grade 3 in MMT) \\
4 & 51&  Left & Mild, able to move the arm against gravity and resistance (Grade 4 in MMT)\\
5 & 66&  Left & Moderate, able to move the arm against gravity (Grade 3 in MMT)\\
6 & 66&  Right & Mild, able to move the arm against gravity and resistance (Grade 4 in MMT)\\
7 & 52&  Right & Moderate, able to move the arm against gravity (Grade 3 in MMT)\\
8 & 60&  Right & Mild, able to move the arm against gravity and resistance (Grade 4 in MMT)\\
9 & 22& Left & Severe, able to move the arm without gravity (Grade 1 in MMT)\\
10 & 29&  Left & Severe, able to move the arm without gravity (Grade 1 in MMT)\\
11 & 50&  Left & Severe, able to move the arm without gravity (Grade 1 in MMT)\\
\hline
\end{tabular}

\end{table*}

We recruited 14 patients from the local hospital, and three of them discontinued their participation during the study because they were uncomfortable with VR devices (Reported dizzy or tired). Consequently, our final sample consisted of 11 patients with upper limb motor impairments, with an average age of 50.5. Our recruitment criteria were based on the extensor and flexor muscles achieving grades 1-4 based on manual muscle testing (MMT) \cite{cuthbert2007reliability} on hands. Participants in these states have a small amount of athletic ability on their hands or lack of endurance. Before official recruitment, we assessed their mental state using the Minimum Mental State Examination (MMSE) \cite{kurlowicz1999mini}, and all participants demonstrated normal cognitive abilities to express emotion and perform simple calculations and got grades higher than 22 in MMSE. None of the participants had prior experience with VR. Approval was obtained from the ethical committees of our home institution and the hospital. All participants gave us written informed consent and received a \$10 shopping card for compensation.
Detailed demographics are shown in Table. 1.

\subsection{Design}
We designed a within-subject study with the following independent variables: 1) \textbf{Hand redirection technique}. it has five levels: No hand redirection, No hand redirection with reduced difficulty, Pre-offset, Scaling offset, and Post-offset. Three different hand redirection techniques are investigated in the study. The parameters for the hand redirection techniques are configured such that the displacement between the virtual and the physical hands is the same across all offset conditions when the users reach maximum reach in the designated direction.  \textit{No hand redirection} is used as a baseline. However, hand redirection effectively reduces the task difficulty by reducing the required travel distance of the user's hand. To understand the effect of hand redirection with tasks of the same difficulty, we included a second baseline - \textit{no hand redirection with reduced difficulty}, in which the target is set closer to the user's hand such that its task difficulty is the same as the three offset conditions.
    % \item \textit{Hand redirection displacement (0.25, 0.5)}: the amount of displacement that results from hand redirection when the user's arm is fully extended in the designated direction in proportion to the maximum reach distance of the affected hand. The displacements are selected based on preliminary tests with healthy participants.
2) \textbf{Hand movement direction}. It has three levels: horizontal, vertical, and depth. The direction in which the target shows up for the participant to reach.
    % \item \textit{Target distance} \textit{(0.5, 0.75, 1.0, 1.25, 1.5)}: the distance between the target and the initial position of the participant's hand in proportion to the maximum reach distance of the affected hand. While target distances 1.0, 1.25, and 1.5 are used for all hand redirection techniques and the baseline, target distances 0.5, and 0.75 are only used with the baseline (no hand redirection). This is because hand redirection effectively reduces the required travel distance of the affected hand thus reducing the task difficulty. To understand the effect of hand redirection with tasks of the same level of difficulty, we introduced two additional shorter target distances such that the baseline's task difficulty is calibrated to the same level as the hand redirection conditions.

% Before our formal study, we conducted preliminary tests with healthy participants to evaluate our system. Participants were asked to reach targets with hand redirection applied to their hands.  We applied scaling or remapping with factors ranging from 1 to 1.5, using a step size of 0.05. Our findings indicated that participants noticed deviations in certain conditions when the remapping factor was 1.25. However, when the remapping factor was increased to 1.5, all participants consistently noticed deviations in all conditions, and they reported that their virtual hands appeared offset or stretched.
Our dependent variables include 1) \textbf{Training efforts}. We used the average hand travel distance in each trial to measure it. We assume that more effect is spent on training when the affected hand reaches further toward the target direction. To account for differences in each participant's ability,  we normalized the hand travel distance by dividing it by its maximum reach distance recorded during the preparation stage.; 2) \textbf{Sense of hand ownership}. This is measured by a questionnaire in which participants reported their level of avatar embodiment on hands \cite{gonzalez2018avatar} on a 7-point Likert scale.

%% overshooting 去掉字数应该下5000，可以解释但是不够strong
% Overshooting (the hand extending beyond the target) can result from losing the semantic sensation of the hand and it is related to the decrease of hand ownership \cite{brozzoli2012s}. As a result, we used the overshooting rate as a metric to measure hand ownership. The overshooting rate is the proportion of trials in which the virtual hand travels beyond the target distance. The overshooting rate can help us understand patients' hand ownership and avatar embodiment in different conditions.

\subsection{Procedure}
\textbf{Preparation and warm-up.} Before the study, participants were told that the study investigated the efficacy of rehabilitation in VR. They are not informed of the real research target.
After signing the informed consent form, participants were assisted in donning the equipment and positioned in front of the BURT Rehabilitation Robot. Their affected hand was securely affixed to the apparatus, as shown in (Fig. \ref{apparauts} A).
Participants found themselves in a laboratory setting inside the VR environment with an empty table before them. They could see their affected hand resting on the table. Participants were informed that their mission was to reach the red ball, and if its color changed to black, they would consider it a success and should return their hand to the initial position.
In cases where participants couldn't accomplish the mission, they were instructed to return their hands to the initial position and wait for another refresh. Each participant was given one minute to explore the virtual environment to get familiar with hand movement and the task. During this warm-up section, no hand redirection techniques were applied.
% \begin{figure}
% \centering
% \includegraphics[scale=0.4]{supplements/Procedure.pdf}
% \caption{The study procedure: Record initial position (A), Record max position on three axes (B), Reach target in a condition (C), and Accomplish questionnaire (D)}
% \label{procedure}   
% \end{figure}

\textbf{Rehabilitation exercise.} Participants were instructed to adjust their hands and body to a comfortable position (Fig. \ref{apparauts} B).  Following this, participants were directed to perform specific movements involving their hands' reaching, flexion, and horizontal abduction to the fullest extent possible along the x, y, and z axes, corresponding to depth, vertical, and horizontal, respectively (Fig. \ref{apparauts} C, D, E). These movements are commonly used for rehabilitation \cite{ellis2018progressive}. The maximum reach distances in these three directions were stored.
Moving forward, a ball was generated relative to the patient's hand's initial position. This target was positioned along the x, y, and z axes, with distances set at 1, 1.25, and 1.5 times the patient's maximum achievable range of motion (calculated from the Max Position to the Initial Position). For each of the three distances, participants completed one action under 1.25 times offset on the x-axis, then three reaching actions under 1.5 times offset. The target refreshes six times on each axis if it is a baseline condition. 
Each patient experienced five conditions in a randomized order for counterbalancing.
The patient will rest and take up the VR headset between each condition to complete the questionnaire and answer the Likert scale questions. This exercise took around 30 minutes.

\textbf{Semi-structured interview.} We conducted a semi-structured interview with nine participants. Two of the 11 participants refused the interview due to personal reasons. During this interview session, we provided participants with information about our genuine research objectives and demonstrated the practical application of hand redirection within the VR environment. We engaged them in discussions about the effects of their efforts, motivation, and sense of hand ownership in VR-based rehabilitation.
Each interview session lasted approximately 20 minutes for each participant.

\subsection{Data Analysis}

For each participant we obtained 90 trials (3 target distance$\times$2 coefficient$\times$3 direction$\times$5 technique), and 11 participants achieved 990 trials in total.
We pre-processed the data by eliminating invalid trails caused by inaccurate hand-tracking (The hands are not detected) and participants could not move their hands due to severe motor impairments in the designated direction (Sum up to 60 trials). 
The Shapiro-Wilk normality test was executed upon eliminating the invalid trails, revealing that the data from questionnaires did not follow a normal distribution. Still, data from travel distances of hands followed a normal distribution. Consequently, the Friedman test \cite{friedman1940comparison} was conducted to assess the difference between data in questionnaires and two-way repeated-measures ANOVA was conducted to assess the difference between data in travel distances.

\section{RESULTS AND FINDINGS}
In this section, we will combine the results of data from participants and open coding interviews to investigate how hand redirection affects people with motor impairments in effort and performance, motivation, and ownership.
\subsection{Effort and Performance}

% Figure \ref{realhand} shows xxx. Then, report key inside in the figure. RM-ANOVA results show xxxx. then explain what the statistical signficant means. 

Questionnaire results show no significant differences between different hand redirection techniques on self-reported performance (p = 0.818) and effort (p = 0.115). 
%%从这里继续

\begin{figure*}[!h]
\centering
\includegraphics[width=\linewidth]{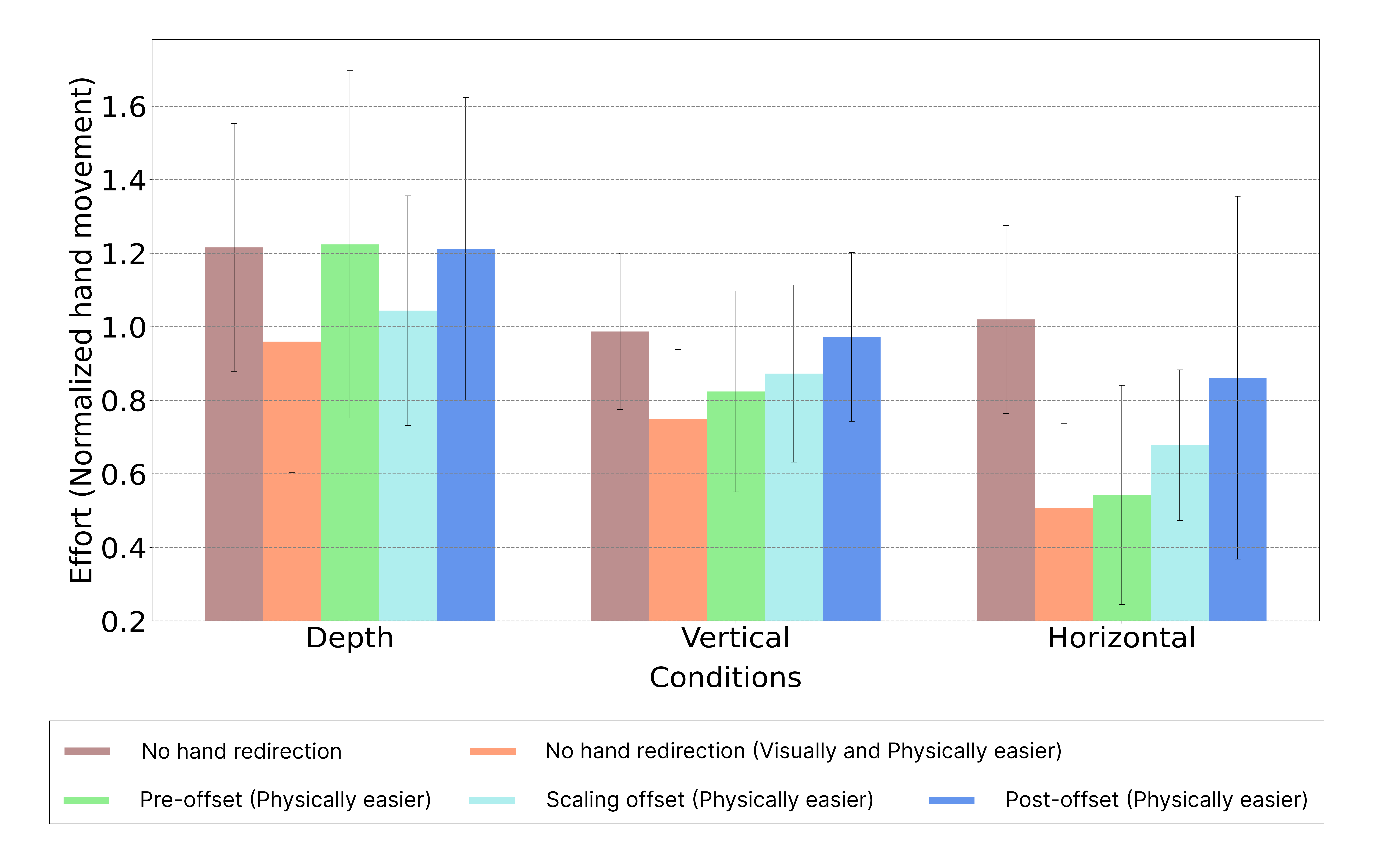}
\caption{The training effort in three movement directions (depth, vertical, and horizontal) with five different hand redirection conditions, including three hand redirection techniques (Pre-offset, Scaling offset, and Post-offset) and two baseline conditions (no hand redirection w/o reduced difficulty). }
\Description{The bar chart summarises normalized effort for each direction and conditions for all participants. On the depth direction, participants have the value of (1.26, 0.96, 1.22, 1.04, 1.21)  on the no hand redirection w/o reduced difficulty condition,  Pre-offset condition, Scaling offset condition and Post-offset condition, respectively. On the vertical direction, participants have the value of (0.99, 0.75, 0.82, 0.87, 0.97,) the no hand redirection w/o reduced difficulty condition,  Pre-offset condition, Scaling offset condition and Post-offset condition, respectively. On the horizontal condition, participants have the value of (1.02, 0.51, 0.54,  0.69, 0.86) on the no hand redirection w/o reduced difficulty condition,  Pre-offset condition, Scaling offset condition and Post-offset condition, respectively.}
\label{realhand}
\end{figure*}

Figure \ref{realhand} shows the effect of different hand direction techniques on effort measured by the hand travel distance. Two-way repeated-measures ANOVA results show a significant difference in the participants' effort to reach the target between different hand redirection techniques (p<.001) and movement directions (p<.001). Among all conditions, participants spent significantly higher efforts in offset conditions (\textit{Pre-offset}, \textit{Scaling offset}, and \textit{Post offsets}) comparing to the baseline  (\textit{No hand redirection with Reduced difficulty}), which means hand redirection can increase patients' effort in reaching tasks during rehabilitation exercise when the task difficulty is the same. Among the three techniques, Post-offset yields the highest increase in effort.
% We identify three key insights. First, participants moved their hands less in the Pre-offset and Scaling offset condition than in the no hand redirection condition. Second, participants' hands reached farther in different directions with hand redirection in the same difficult level, especially in the Post-offset condition. Third, participants showed more effort in the depth direction.

\subsection{Motivation}
During the interview, almost all participants thought that hand redirection could enhance their motivation, as it helped them complete the task. ( \emph{P4: "I feel frustrated when I can not reach the target, so I want to be helped."
P5: "It is better to deploy hand redirection, which can improve the range of movement. 
If I can not reach the target, I will give up. This can improve my confidence because my confidence comes from completing tasks. If I perform worse, he will lose confidence and motivation. For long-term rehabilitation, this will also improve my motivation. "})
Only one participant (P10) rejected it as he preferred to see his truthful capability in the virtual scene during exercises.

When asked whether the task difficulty should be reduced by placing targets closer to the affected hand so that they can reach more targets, most participants objected. Some participants believe that closer targets will make the exercise boring (\emph{P4:" It is boring when the target is too near to the hand.", P5:" It seems too easy to reach, and makes it boring, losing interest."}). Other participants worried that easier tasks may hinder the therapeutic effect.\textit{ (P6: "The task should be more difficult because I exercise here. Too easy may bring no effect on rehabilitation)}
\subsection{Sense of Hand Ownership}
Results from interviews and questionnaires show that participants hardly notice hand redirection, which implies a high sense of hand ownership. 
RM-ANOVA test shows no significant differences between the five conditions on self-reported hand movement consistency (p = 0.39), hand location consistency (p = 0.745), and hand ownership (p = 0.123). 
During the interview, when asked whether they noticed there was displacement between the real hand and the virtual hand, only two participants said they did. After revealing our true research objectives and demonstrating different hand redirection techniques, two more participants (P4, P6) said that they noticed the displacement but they did not report it as they were unable to articulate their observation so they chose to disregard them (\emph{P6: "There are differences in my hands, But I can not tell what exactly the differences are. The hands in the VR seem less flexible."}). 

% Another participant mentioned that they felt they couldn't reach the target, leading them to perceive that the virtual hands did not belong to them(P4, P6, P9). Notably, these participants believed they could reach all targets before the study began.
% \begin{figure}
% \centering
% \includegraphics[scale=1.5]{supplements/overshooting.pdf}
% \caption{The overshooting rate among techniques and three directions.}
% \Description{The bar chart summarises the overshooting rate for each direction and conditions for all participants. On the no hand redirection condition, participants have the value of (1.26, 0.99, 1.02) and on depth, vertical and horizontal direction, respectively. On the no hand redirection with reduced difficulty condition, participants have the value of (0.96, 0.75, 0.51) on depth, vertical and horizontal direction, respectively. On the Pre-offset condition, participants have the value of (1.22, 0.82, 0.54) on depth, vertical and horizontal direction, respectively. On the Scaling offset condition, participants have the value of (1.04, 0.87, 0.69) on depth, vertical and horizontal direction, respectively. On the Post-offset condition, participants have the value of (1.21, 0.97, 0.86) on depth, vertical and horizontal direction, respectively.}
% \label{virtualhand}
% \end{figure}

% ANOVA test shows significant differences in overshooting rate among different hand redirection techniques(p<.1) and directions(p<.01) , as shown in Fig. \ref{virtualhand}. Participants always overshoot in the vertical direction, which corresponds to shoulder flexion.

\subsection{Preference on Hand Redirection Techniques}

Interview results show most participants prefer to use the Post-offset for future VR-based rehabilitation exercises. During the interview, participants were asked to choose a technique for future rehabilitation in VR. While two participants have no preference (\emph{P4, P8: "All methods are great because I can reach more targets. I can accept hand does not belong to me in VR rehabilitation"}), other participants had the following preferences:

\textbf{Post-offset.}
Most participants prefer the Post-offset for two different reasons: 1) Post-offsets could enhance their endurance \emph{(P3: "I want to be helped at the end when I can not reach the target."  The pleasure lies in the accomplishment achieved through diligent effort", P6: "In Post-offset condition, the hand is reaching to target slowly. It'll build up my endurance." }) 2) It encourages the participant to try harder. (\textit{P5: "Help me at the end is a better technique because there is an expected value that I will not easily give up. As for the other two redirection methods, the task seems easier when offset appears since the beginning of each trial. Then, I will not try harder and may give up.})

\textbf{Scaling offset.}
P7 preferred Scaling offset as it is more entertaining  (\emph{P7: "The Post-offset is similar to Scaling offset, so it is hard to say which is better. But I will choose Scaling offset because it is much more exciting than other techniques."}).

\textbf{Pre-offset.}
P9 preferred Pre-offset due to its low noticeability. However, other participants suggested that once it is noticed, the Pre-offset felt unrealistic (\emph{P6: "It is disproportionate." P7:" It is not realistic. Moreover, it is much more excessive than Scaling offset."}).

\textbf{Additional comments on the systems.}
The system should incorporate haptic feedback.
(\emph{P5: "The environment is great and real, but the appearance of the hands is not real enough. The haptic of hands is also lacking. When I can not feel haptic from my disabled hand, I try another hand, but I still can not feel anything.})
We also find this phenomenon in other participants. They will try their other hand to seek haptic feedback. This participant also gave us feedback compared with other devices in daily rehabilitation in the hospital. 
\emph{I felt as if addicted to it. When using this system, it is not easy to make me feel tired. This is far more interesting than other devices I experienced in daily rehabilitation"} 
Young people may be less likely to accept hand redirection.
(\emph{P6: "The young patients may not easily accept this method. They are sensitive to the help from the machine."}) But the P9 disagrees with that. (\emph{P9: "I played so many computer games so I can notice the difference between this VR training and other training. I am in a game. I do not care if I can push my hand to limit this way.})

\section{DISCUSSION AND FUTURE WORK}
We investigated how hand redirection affects rehabilitation for people with upper-limb motor impairments. In summary, our findings show that hand redirection improves the training effort without losing  the sense of hand ownership for people with motor impairments. Furthermore, almost all participants expressed a strong inclination toward incorporating hand redirection. They believed it could enhance their confidence in the rehabilitation process by completing more missions, and they will be more motivated \cite{toure2014measure}. Among all hand redirection conditions, the proposed Post-offset enhanced participants' effort and performance and kept more ownership, while participants felt it to be more motivating.  
%We know that the participants show less semantic sensation in their hands. As a result, we can design a higher value, change the implementation time for the Post-offset, and make participants perform better with higher ownership in future work.
% We made the following key contributions: a) a novel hand redirection technique designed for VR-based upper limb rehabilitation. b) an empirical study that investigates how hand redirection techniques affect the effort, motivation, and sense of hand ownership of people with upper limb motor impairments in VR-based rehabilitation. 
Below we discussed our findings.

\subsection{Training Effort}
\textbf{Visually more difficult tasks lead to higher efforts}.
% Hand redirection improves efforts when patients are experiencing higher difficult tasks
Participants reach farther with hand redirection than the baseline condition with reduced difficulty. 
This might be the result of the smaller visual target distance in the baseline with reduced difficulty which makes participants subconsciously put in less effort.

\textbf{Reduced task difficulty resulting from hand redirection can lead to lower efforts}.
It is more difficult for participants to reach the target in \textit{no hand redirection }condition compared to offset conditions as hand redirection effectively reduces the required travel distance of the hand for reaching the target. Our result shows that participants spend significantly less effort in  \textit{Pre-offset} and \textit{Scaling offset}. \textit{Post-offset} also exhibits less training effort compared to \textit{no hand redirection} in two directions though there is no statistical significance. These results show that reduced task difficulty results from hand redirection can lead to lower efforts. 

\textbf{Post-offset outperforms Pre-offsets and Scaling offsets in retaining training effort}.
Participants spent significantly more effort in rehabilitation exercises with Post-offset compared to other offset conditions. The rationale behind this was revealed in the interview in which participants suggested that\textit{ Post-offset} only helps them at the end when they cannot reach the target, which can be helpful in building up their endurance. These results show that Post-offset has the potential to reduce task difficulty while avoiding slacking, which could decrease patients' motor output and decrease recovery \cite{casadio2012learning}.

\textbf{Participants reach farther in the depth direction}.
% Among no hand redirection condition and hand redirection conditions, the targets are shown at the same position as the participants. The difference is that participants move less in the hand redirection conditions but can still reach the targets. 
This is because the participant can lean forward their body to compensate for their impairment when reaching the target. Compensation is common because people with motor impairment, as it is necessary for completing most activities of daily living \cite{jones2017motor}.

\subsection{Motivation}
\textbf{Participants believe hand redirection can enhance their motivation in these rehabilitation tasks}
According to our results from semi-structured interviews, almost all participants in this study reported they would like to use hand redirection in future VR rehabilitation. They believe it could enhance their motivation and confidence in the long term due to more tasks they can achieve, which is also reflected by the results of the efforts. 
In future work, we can conduct a longitudinal rehabilitation study to evaluate how hand redirection motivates patients in the long term. We can also combine the evaluation results in the hospital, and compare the training effect between VR-based rehabilitation with hand redirection and traditional training methods.

% \revision{\textbf{Participants prefer tasks with higher difficulty to obtain more progress}
% The comfort zone is usually discussed in scenes of social context, including education\cite{lara2023challenge}, occupation\cite{stephenson2018exploration}, or individual development\cite{leberman2002does}, but seldom in rehabilitation. Participants in our study talked about how they want to break through, which is also our initial intention, to help them see themselves pushing the limits. In future work, this can help define VR rehabilitation for people with motor impairments, to design rehabilitation programs of reasonable difficulty to maximize patient motivation.
% }

\textbf{Participants object to reducing task difficulty in an obvious manner} Although participants consider hand redirection techniques (which reduce task difficulty) helpful, most participants ($N=10$)  objected to reducing task difficulty by simply placing the target closer as they think it makes the exercise less challenging and may harm the therapeutic effect. This suggests that patients may prefer tasks that are visually more challenging during rehabilitation exercises. This may be explained by the comfort zone learning model, which theorizes that learners learn better when tasks are more difficult and outside of their comfort zone \cite{leberman2002does}. Future work could investigate how visual difficulty (visually challenging or complex tasks \cite{ragan2015effects}) affects the motivation and effectiveness of rehabilitation.

\subsection{Virtual Hand Ownership}
\textbf{Participants hardly notice hand redirection}.
Most participants could not notice the difference between different conditions and gave full points for a sense of hand ownership in questionnaires.
This finding was corroborated by the results of semi-structured interviews, which show most participants are not aware of the disparities between the virtual hand and the real hand. In future work, we may use quantitative methods such as EEG \cite{feick2023investigating} or EMG \cite{salagean2022virtual} to investigate how hand redirection affects patients' sense of hand ownership and help explore the optimal hand redirection threshold in people with motor impairments.

\section{LIMITATION}
% In this work, three main directions can be further investigated: long-term usage, personalization, and novel-designed serious games.
While our results suggest promise of VR hand redirection for rehabilitation, this work is limited in the following aspects.
First, medical evaluation is not included in the study. Due to limited time and resources, each participant only experienced the VR-based rehabilitation application for less than one hour which is too short for any noticeable therapeutic outcomes.  Future work can investigate the long-term effect of hand redirection on therapeutic outcomes.
Second, our results may not apply to patients with mild symptoms as the study was conducted with hospitalized patients with upper limb muscle strength levels between 1 to 4, while people with higher muscle strength might exhibit less sense of hand ownership due to better proprioception. 
Third, the warm-up section lasts for only 1 minute. This may cause the participants to remain unfamiliar with VR and hand redirection and make it harder to find virtual hand displacements.
Fourth, VR equipment and app interface can also impact patient motivation and embodiment in VR due to insufficiently realistic environments and unfamiliar operation methods. This could be investigated through questionnaires in future work. 
Finally, although we have selected reasonable parameters (e.g. target distances, displacement for hand redirection) for VR-based rehabilitation exercises with hand redirection by testing with abled-bodied people, these parameters are not optimal. Future work can investigate the optimal value of displacement for hand redirection or target distances in VR-based rehabilitation in general.

% We did not investigate the noticeability threshold of redirection techniques for people with motor impairment. Their sensation of body ownership may recover back to normal. Therefore, progressively adjustable hand redirection should be considered.

% Moreover, we didn't include other variables in the study, such as gender, age, personal ability, and social context. The influence of these factors on patients' perception of hand redirection may warrant further investigation, allowing for more personalized rehabilitation exercises. 

% Further, future work can explore the combination of VR hand redirection and actuated interface to provide haptic feedback for a more immersive and engaging experience during rehabilitation exercises. 

% In future work, we can use this characterization to design novel, serious games for rehabilitation. On one hand, it can reduce the likelihood of decreased compliance due to the lack of interest in rehabilitation games. On the other hand, it allows patients to experience more engaging hand redirection and customization games that they are difficult to lose a sense of body ownership.

\section{CONCLUSION}
% With continuous rehabilitation, people with motor impairments acquire more dramatic recovery results. However, the limited effectiveness, the heavy training burden, and resulting fatigue can lead to decreased motivation and confidence, consequently fostering resistance to rehabilitation training.
% Hand redirection in VR has the potential to help participants see significant performance in rehabilitation. Previous work tested thresholds of hand redirection in VR, but those mainly focused on healthy participants instead of people with motor impairments. 
Hand redirection holds immense potential in improving the motivation of people with upper limb motor impairment during VR-based rehabilitation exercises. While prior research in hand redirection mostly focuses on the experience of hand redirection for abled-bodied people, this study examined the effect of various hand redirection techniques on the motivation, effort, and sense of hand ownership of people with upper limb motor impairment during VR-based rehabilitation exercises.  We proposed a novel hand redirection technique - \textit{Post offset} for upper limb rehabilitation. A user study was conducted at a local hospital with  11 participants with upper limb motor impairment. Our findings suggest that while people with upper limb motor impairment can hardly notice hand redirection, hand redirection improves their motivation and effort during VR rehabilitation exercises. Participants show the highest effort with Post-offset among all hand redirection techniques investigated in this study. Most participants express interest in incorporating hand redirection into future long-term VR rehabilitation programs. 
\begin{acks}
This work is partially supported by 1) 2024 Guangzhou Science and Technology Program City-University Joint Funding Project (PI: Mingming Fan); 2) 2023 Guangzhou Science and Technology Program City-University Joint Funding Project (No. 2023A03J0001) ; 3) Guangdong Provincial Key Lab of Integrated Communication, Sensing and Computation for Ubiquitous Internet of Things (No.2023B1212010007).
\end{acks}
\bibliographystyle{ACM-Reference-Format}
\bibliography{main_bib}

\end{document}